\documentclass[aps,twocolumn,showpacs,nofootinbib,showkeys,superscriptaddress,preprintnumbers,longbibliography,10pt]{revtex4-1}
\usepackage[utf8]{inputenc}

\usepackage{latexsym}
\usepackage{epsfig}
\usepackage{graphicx,subfigure}
\usepackage[normalem]{ulem}
\usepackage{color}
\usepackage{xcolor}
\usepackage{multirow}
\usepackage{hyperref}
\usepackage{numprint}
\usepackage{amsmath}
\usepackage{amssymb}

\usepackage{xcolor}

\begin{document}

\def\ga{\mathrel{\raise.3ex\hbox{$>$\kern-.75em\lower1ex\hbox{$\sim$}}}}
\def\la{\mathrel{\raise.3ex\hbox{$<$\kern-.75em\lower1ex\hbox{$\sim$}}}}

\def\be{\begin{equation}}
\def\ee{\end{equation}}
\def\bea{\begin{eqnarray}}
\def\eea{\end{eqnarray}}

\def\betap{\tilde\beta}
\def\del{\delta_{\rm PBH}^{\rm local}}
\def\Msun{M_\odot}

\newcommand{\dd}{\mathrm{d}} 
\newcommand{\Mpl}{M_P} 
\newcommand{\mpl}{m_\mathrm{pl}} 

\newcommand{\CHECK}[1]{{\color{red}~\textsf{#1}}}
\newcommand{\new}[1]{{\color{blue} #1}}
 
\newcommand{\SNRmfPv}{$7.98^{+0.62}_{-1.03}$}
\newcommand{\SNRmfHPv}{$6.38^{+0.52}_{-1.31}$}
\newcommand{\SNRmfLPv}{$5.21^{+0.54}_{-0.87}$}
\newcommand{\PrimMassPv}{$4.65^{+1.21}_{-2.15}$}
\newcommand{\SecMassPv}{$0.77^{+0.50}_{-0.12}$}
\newcommand{\PrimSpinPv}{$0.32^{+0.47}_{-0.26}$}
\newcommand{\SecSpinPv}{$0.48^{+0.46}_{-0.43}$}
\newcommand{\qPv}{$0.17^{+0.34}_{-0.05}$}
\newcommand{\MtotPv}{$5.42^{+1.10}_{-1.65}$}
\newcommand{\ChieffPv}{$-0.06^{+0.17}_{-0.32}$}
\newcommand{\ChipPv}{$0.28^{+0.34}_{-0.21}$}
\newcommand{\DLPv}{$119^{+82}_{-48}$}
\newcommand{\zPv}{$0.028^{+0.018}_{-0.010}$}
\newcommand{\raPv}{$-2^{+34}_{-35}$}
\newcommand{\decPv}{$47^{+14}_{-26}$}
\newcommand{\MfPv}{$5.34^{+1.11}_{-1.70}$}
\newcommand{\SfPv}{$0.39^{+0.24}_{-0.07}$}
\newcommand{\ProbSSMPv}{$85\%$}
\newcommand{\MaxSNRmfPv}{$9.09$}
\newcommand{\ProbMGPv}{$64\%$}

\newcommand{\SNRmfXPHM}{$7.94^{+0.70}_{-1.05}$}
\newcommand{\PrimMassXPHM}{$4.71^{+1.57}_{-2.18}$}
\newcommand{\SecMassXPHM}{$0.76^{+0.50}_{-0.14}$}
\newcommand{\PrimSpinXPHM}{$0.36^{+0.46}_{-0.30}$}
\newcommand{\SecSpinXPHM}{$0.47^{+0.46}_{-0.42}$}
\newcommand{\qXPHM}{$0.16^{+0.34}_{-0.06}$}
\newcommand{\MtotXPHM}{$5.47^{+1.43}_{-1.68}$}
\newcommand{\ChieffXPHM}{$-0.05^{+0.22}_{-0.35}$}
\newcommand{\ChipXPHM}{$0.33^{+0.33}_{-0.26}$}
\newcommand{\DLXPHM}{$124^{+82}_{-48}$}
\newcommand{\zXPHM}{$0.028^{+0.017}_{-0.011}$}
\newcommand{\raXPHM}{$-1^{+34}_{-37}$}
\newcommand{\decXPHM}{$46^{+14}_{-29}$}
\newcommand{\MfXPHM}{$5.40^{+1.45}_{-1.73}$}
\newcommand{\SfXPHM}{$0.42^{+0.22}_{-0.10}$}
\newcommand{\ProbSSMXPHM}{$84\%$}
\newcommand{\MaxSNRmfXPHM}{$9.18$ }
\newcommand{\ProbMGXPHM}{$59\%$}

\newcommand{\lnBHL}{$7.00 \pm 0.10$}
\newcommand{\lnBH}{$1.56\pm 0.07$}
\newcommand{\lnBL}{$0.48 \pm 0.06$}
\newcommand{\lnBcohinc}{$4.96 \pm 0.13$}

\title{{Analysis of a subsolar-mass compact binary candidate} \\ from the second observing run of Advanced LIGO}

\author{Gonzalo Morr\'as}
\affiliation{Instituto de F\'isica Te\'orica UAM/CSIC, Universidad Aut\'onoma de Madrid, Cantoblanco 28049 Madrid, Spain}

\author{Jos\'e Francisco Nu\~no Siles}
\affiliation{Instituto de F\'isica Te\'orica UAM/CSIC, Universidad Aut\'onoma de Madrid, Cantoblanco 28049 Madrid, Spain}

\author{{Juan Garc\'ia-Bellido}}
\affiliation{Instituto de F\'isica Te\'orica UAM/CSIC, Universidad Aut\'onoma de Madrid, Cantoblanco 28049 Madrid, Spain}

\author{{Ester Ruiz Morales}}
\affiliation{Departamento de F\'isica, ETSIDI, Universidad Polit\'ecnica de Madrid, 28012 Madrid, Spain}
\affiliation{Instituto de F\'isica Te\'orica UAM/CSIC, Universidad Aut\'onoma de Madrid, Cantoblanco 28049 Madrid, Spain}

\author{Alexis Men\'endez-V\'azquez}
\affiliation{Institut de F\'\i sica  d'Altes Energies (IFAE), Barcelona Institute of Science and Technology, E-08193 Barcelona, Spain}

\author{Christos Karathanasis}
\affiliation{Institut de F\'\i sica  d'Altes Energies (IFAE), Barcelona Institute of Science and Technology, E-08193 Barcelona, Spain}

\author{Katarina Martinovic}
\affiliation{Theoretical Particle Physics and Cosmology Group, \, Physics \, Department, King's College London, \, University \, of London, \, Strand, \, London \, WC2R \, 2LS, \, UK}

\author{Khun Sang Phukon}
\affiliation{Nikhef - National Institute for Subatomic Physics, Science Park, 1098 XG Amsterdam, The Netherlands}
\affiliation{Institute for High-Energy Physics, University of Amsterdam, Science Park, 1098 XG Amsterdam, The Netherlands}
\affiliation{Institute for Gravitational and Subatomic Physics, Utrecht University, Princetonplein 1, 3584 CC Utrecht, The Netherlands}
\affiliation{School of Physics and Astronomy and Institute for Gravitational Wave Astronomy, University of Birmingham, Edgbaston, Birmingham, B15 9TT, United Kingdom}

\author{Sebastien Clesse}
\affiliation{Service de Physique Th\'eorique, Universit\'e Libre de Bruxelles (ULB), Boulevard du Triomphe, CP225, B-1050 Brussels, Belgium}

\author{Mario Mart\'\i nez}
\affiliation{Institut de F\'\i sica  d'Altes Energies (IFAE), Barcelona Institute of Science and Technology, E-08193 Barcelona, Spain}
\affiliation{Instituci\'o Catalana de Recerca i Estudis Avançats (ICREA), Barcelona, Spain}

\author{Mairi Sakellariadou}
\affiliation{Theoretical Particle Physics and Cosmology Group, \, Physics \, Department, King's College London, \, University \, of London, \, Strand, \, London \, WC2R \, 2LS, \, UK}
\date{\today}
\begin{abstract} 

We perform an exhaustive follow-up analysis of a subsolar-mass (SSM) gravitational wave (GW) candidate reported by Phukon et al. from the second observing run of Advanced LIGO. {This candidate has a reported} signal-to-noise ratio (SNR) of $8.6$ and false alarm rate of $0.41$ yr which are too low to claim a clear gravitational-wave origin. When improving on the search by using more accurate waveforms, extending the frequency range from 45 Hz down to 20 Hz, and removing a prominent blip glitch, we find that the posterior distribution of the network SNR lies mostly below the search value, with the 90\% confidence interval being \SNRmfXPHM. Assuming that the origin of the signal is a compact binary coalescence (CBC), the secondary component is $m_2 = \text{\SecMassXPHM} M_\odot$, with $m_2 < 1 M_\odot$ at \ProbSSMXPHM~confidence level, suggesting an unexpectedly light neutron star or a black hole of primordial or exotic origin. The primary mass would be $m_1 = \text{\PrimMassXPHM} M_\odot$, likely in the hypothesized lower mass gap and the luminosity distance is measured to be $D_{\rm L}=$\DLXPHM Mpc. We then probe the CBC origin hypothesis by performing the signal coherence tests, obtaining a log Bayes factor of \lnBcohinc~for the coherent vs. incoherent hypothesis. We demonstrate the capability of performing a parameter estimation follow-up on real data for an SSM candidate with moderate SNR. The improved sensitivity of O4 and subsequent LIGO-Virgo-KAGRA observing runs could make it possible to observe similar signals, if present, with a higher SNR and more precise measurement of the parameters of the binary.

\end{abstract}

\maketitle

\section{Introduction}

The development of gravitational wave astronomy, with about 90 {compact binary} coalescence (CBC) events detected so far~\cite{Abbott:2016blz,TheLIGOScientific:2016pea,LIGOScientific:2018mvr,LIGOScientific:2020ibl,LIGOScientific:2021usb,LIGOScientific:2021djp} by the LIGO-Virgo-KAGRA (LVK) collaboration~\cite{LVKarticle}, is driving a true revolution in astrophysics and cosmology. As the number of detected events grows with successive observing catalogs, { the range of the inferred component masses has extended to previously unexplored regions,} with black holes (BH) found~\cite{Abbott:2020tfl,Abbott:2020mjq} in the pair-instability mass gap~\cite{Woosley:2016hmi} and in the hypothesized lower mass gap~\cite{Abbott:2020khf}.
The frequency range of the LIGO~\cite{AdvancedLigo} and Virgo~\cite{AdvancedVirgo} detectors makes them also sensitive to CBC signals in which one of the compact objects has a mass below 1$M_\odot$. 
The detection of a sub-solar mass (SSM) compact object would be of utmost interest since it would require either modification of the standard astrophysical evolution and collapse of ordinary matter or a new formation mechanism operating in the Universe, such as primordial black holes (PBHs)~\cite{1967SvA....10..602Z,Hawking:1971ei,Carr:1974nx,1975Natur.253..251C,GarciaBellido:1996qt} or SSM objects originated by dark matter with exotic properties~\cite{Kouvaris:2010jy,deLavallaz:2010wp,Bramante:2014zca,Bramante:2015dfa,Bramante:2017ulk,Kouvaris:2018wnh,Shandera:2018xkn,Chang:2018bgx,Latif:2018kqv,Dasgupta:2020mqg,Gurian:2022nbx,Ryan:2022hku,Hippert:2022snq}.

Before the advent of GW astronomy, the only way to detect SSM black holes was via X-ray binaries~\cite{Corral-Santana:2015fud} or microlensing~\cite{Paczynski:1985jf}. At present, some hints of the existence of such light black holes come from microlensing events towards the bulge~\cite{2020A&A...636A..20W}, from Andromeda~\cite{Niikura:2017zjd} and lensed quasars~\cite{Hawkins:2020rqu,Hawkins:2022vqo}, although the mass and the abundance of the lenses remain uncertain.
Complementary to these astrophysical searches, GW signals of CBC with at least one subsolar component have been searched for in the first (O1), second (O2) and third (O3) observing runs of LVK~\cite{LIGOScientific:2018glc,LIGOScientific:2019kan,Nitz:2021mzz,LIGOScientific:2021job,LIGOScientific:2022hai}, without finding compelling evidence for a clear detection.
Nevertheless, a further search in the O2 data for SSM black holes with low mass ratios ~\cite{Phukon:2021cus} and the latest O3b SSM search results from LVK~\cite{LIGOScientific:2022hai} have reported several potential candidates with a false alarm rate smaller than $2\, \rm{yr^{-1}}$.  

In this work, we follow up on the O2 search reported in \cite{Phukon:2021cus}, using the standard parameter estimation (PE) methods {to further investigate} the candidates reported in Table I. {Given that PE on these long GW signals is extremely time consuming, we have focused on the third candidate of the table, which is the lowest FAR trigger found in coincidence by both LIGO Hanford and LIGO Livingston interferometers,} {which allows more confident rejection against terrestrial noise}. This candidate was observed on April 1st 2017 and we will refer {to} it here as SSM170401. 

{In this analysis we have extended the frequency range of the search from 45 Hz down to 20 Hz. We have also improved upon the \texttt{TaylorF2}~\cite{Buonanno:2009zt} waveform used in the template bank of search, by using for PE the more accurate waveforms \texttt{IMRPhenomPv2}~\cite{PhysRevLett.113.151101} and \texttt{IMRPhenomXPHM}~\cite{Pratten_2021} that include the merger and ringdown phases, as well as spin precession and, in the case of \texttt{IMRPhenomXPHM}, higher order modes.} We have also inspected the quality of the data and discussed the impact of a prominent glitch removal using standard tools such as \texttt{BayesWave} \cite{Cornish_2015,PhysRevD.91.084034,Davis_2022}. 
Finally, assuming that the {origin} of the {candidate} is a BBH merger, the PE allows us to infer the component masses, spins, distance and sky location, as well as the {posterior} probability of having an SSM component {of the} hypothetical source for SSM170401. 

In the following sections, we describe  in detail different aspects of this analysis that reveal the peculiarities and difficulties of doing PE on this type of candidate, as well as the necessary analysis tools in preparation for a possible future significant candidate, given the increase of sensitivity expected in the O4 run.

\section{Significance of SSM170401}
\label{sec:significance}

The {candidate} was found in data taken on  April 1st, 2017, 01:43:34 UTC during the O2 LIGO-Virgo observing run. {It} was not reported by any of the LVK searches, both generic \cite{LIGOScientific:2018mvr} and SSM specific \cite{abbott2019search}, but it was {found in a dedicated} search for SSM {mergers} in asymmetric binaries using the \texttt{GstLAL} pipeline~\cite{Phukon:2021cus}. The {search reported} detector frame masses of 4.897 $M_\odot$ and 0.7795 $M_\odot$, with a false-alarm-rate (FAR) of 0.4134 yr$^{-1}$ and a combined network signal-to-noise ratio (SNR) of $\sim 8.67$. Given that the time of O2 coincident data suitable for observation is $T_\mathrm{obs} = 118 ~\mathrm{days}$ \cite{LIGOScientific:2018mvr}, the false alarm probability (FAP) of this candidate, according to the search is:

\begin{equation}
    \mathrm{FAP} = 1 - \exp\left\{ - \mathrm{FAR}\cdot T_\mathrm{obs}\right\} = 0.12 \, .
    \label{eq:FAP}
\end{equation}

The interpretation of this FAP is that the search would produce a higher-ranked candidate in 12\% of trials over data containing only noise.

We can also estimate an upper bound for the probability of this signal coming from a CBC merger with an SSM component, given the upper limits on event rates obtained from the O3 SSM searches. \cite{LIGOScientific:2021job,LIGOScientific:2022hai}. In Ref.~\cite{LIGOScientific:2022hai} the 90$\%$ C.L. constraints on the merger rate $\mathcal{R}_{90}$ of SSM binaries are reported in the $(m_1,m_2)$ plane, assuming null results of these searches. For the median values of the component masses of the source of SSM170411 (see Table~\ref{table:parameters}), we find $\mathcal{R}_{90} \sim 2 \times 10^{2}~\mathrm{Gpc}^{-3} \mathrm{yr}^{-1}$. Moreover, in the search where the signal was identified~\cite{Phukon:2021cus}, the volume-time surveyed for these same masses is reported to be $\langle V T \rangle \sim 3 \times 10^{-3}~\mathrm{Gpc}^3 \mathrm{yr} $. Since the arrival of GWs from binary mergers to the detectors is Poisson distributed, with an expected number of events $\mu = \mathcal{R} \langle V T \rangle$, the probability of finding $n$ events would be:
\begin{equation}
    P(n) = \frac{\mu^n}{n!} e^{-\mu} \, .
    \label{eq:Poisson_dist}
\end{equation}
Using the values previously mentioned for $\mathcal{R}_{90}$ and $\langle V T \rangle$, the upper bound on the expected number of events is $\mu_{90} \sim 0.6$ at $90\%$ C.L. and the corresponding upper bound on the probability for the search in Ref.~\cite{Phukon:2021cus} to have found one or more events would be smaller than $1 - P_{90}(0) \sim 0.45$. Therefore, the results of O3 do not particularly constrain the possibility that SSM170401 could come from a real SSM merger.

\begin{figure}[t!]
\begin{center}
\includegraphics[width=0.48\textwidth]{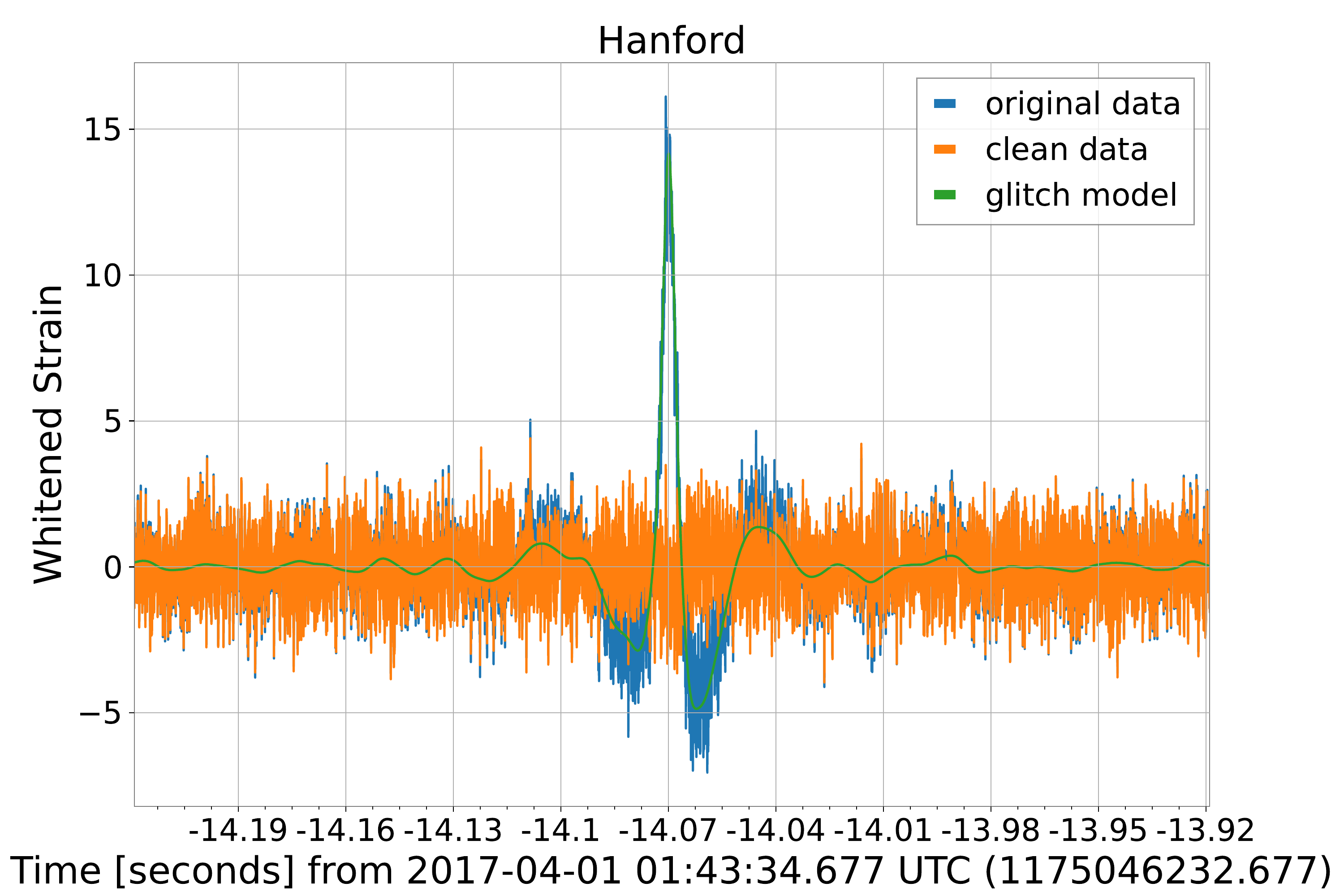}
\end{center} 
\caption{Figure showing the Hanford original whitened strain $\tilde{h}^\mathrm{whitened}(f) = \tilde{h}(f)/\sqrt{S_n(f)}$, the whitened glitch model and the whitened clean data after subtracting the glitch. Times are shown relative to the trigger time.}
\label{fig:H1_glitch}
\end{figure}

The strain in Hanford presents a glitch 14 s before coalescence, as shown in Fig.~\ref{fig:H1_glitch}. The search presented in Ref.~\cite{Phukon:2021cus} uses templates starting at 45~Hz. The loudest template, in this case, is only 10~s long and so should be unaffected by the glitch. However, PE was performed with templates starting at 20~Hz, which are roughly 100~s long for the component masses discussed. In this situation the glitch must be removed. Using \texttt{BayesWave} \cite{Cornish_2015,PhysRevD.91.084034}, we model excess power in the detectors as a sum of sine-Gaussian wavelets. We fit for the glitch and the PSD of the Gaussian noise component simultaneously. We ignore the modelling of the signal due to the extremely low coherent energy per frequency bin deposited in the detectors in the 0.3~s duration of the glitch by such low mass sources, even more when the subtraction is done $\sim 14$~s before coalescence. The same procedure is used routinely by the LVK collaboration in the main GW catalog~\cite{LIGOScientific:2021djp}.

\section{Properties of the source of SSM170401}\label{sec:properties_of_the_source}

To obtain the properties of the {potential source of SSM170401} we interpret the signal as coming from the coalescence of two compact objects. We infer the CBC parameters of the signal using a Bayesian analysis of the data from LIGO Livingston and LIGO Hanford, following the methodology outlined in Appendix B of Ref.~\cite{LIGOScientific:2018mvr}. In analysing the data, we fit two different waveform models: $\texttt{IMRPhenomPv2}$ \cite{PhysRevLett.113.151101} and $\texttt{IMRPhenomXPHM}$ \cite{Pratten_2021}, the latter including higher order modes.  Comparing the PE analyses using the two waveforms models, we find that their posterior distributions are consistent with each other, noting that both of them take into account precessing spins.

\begin{table}[t!]
	\begin{ruledtabular}
	\begin{tabular}{lcr}
	   Parameter & \texttt{IMRPhenomPv2} & \texttt{IMRPhenomXPHM}  \\ \hline\\
	   Signal to Noise Ratio & \SNRmfPv & \SNRmfXPHM \\[4pt]
	   Primary mass ($M_\odot$)   & \PrimMassPv & \PrimMassXPHM \\[4pt] 
	   Secondary mass ($M_\odot$)  & \SecMassPv  & \SecMassXPHM  \\[4pt] 
		Primary spin magnitude  & \PrimSpinPv   &  \PrimSpinXPHM \\[4pt] 
		Secondary spin magnitude  &  \SecSpinPv  &  \SecSpinXPHM    \\[4pt]
		Total mass ($M_\odot$)  & \MtotPv  &  \MtotXPHM  \\[4pt]
		Mass ratio $(m_2/m_1 \leq 1)$  & \qPv  &  \qXPHM \\[4pt]
		$\chi_{\text{eff}}$ \cite{PhysRevLett.106.241101,PhysRevD.82.064016} & \ChieffPv  & \ChieffXPHM \\[4pt] 
		$\chi_{\text{p}}$ \cite{PhysRevD.91.024043}  & \ChipPv  & \ChipXPHM \\[4pt]
		Luminosity Distance (Mpc) & \DLPv & \DLXPHM \\[4pt] 
		Redshift & \zPv & \zXPHM \\[4pt]
		Ra ($^\circ$) & \raPv & \raXPHM \\[4pt] 
		Dec ($^\circ$) &  \decPv   & \decXPHM \\[4pt]
		Final mass ($M_\odot$) & \MfPv  & \MfXPHM \\[4pt] 
		Final spin & \SfPv & \SfXPHM \\[4pt]
		$P( m_2< 1\ M_{\odot})$ & \ProbSSMPv & \ProbSSMXPHM  \\[4pt]
	\end{tabular}
	\end{ruledtabular}
	\caption{Parameters of the source of SSM170401. All masses are in the source frame. We assume $\textit{Planck15}$ Cosmology \cite{Planck15}. The statistical uncertainty of all the parameters is quantified by the equal-tailed $90\%$ credible intervals about the median of the marginalized one-dimensional posteriors. Right ascension (Ra) and declination (Dec) are measured in the International Celestial Reference System (ICRS) \cite{Arias95}. 
	}
	\label{table:parameters}
\end{table}

\begin{figure}[t!]
\begin{center}
\includegraphics[width=0.48\textwidth]{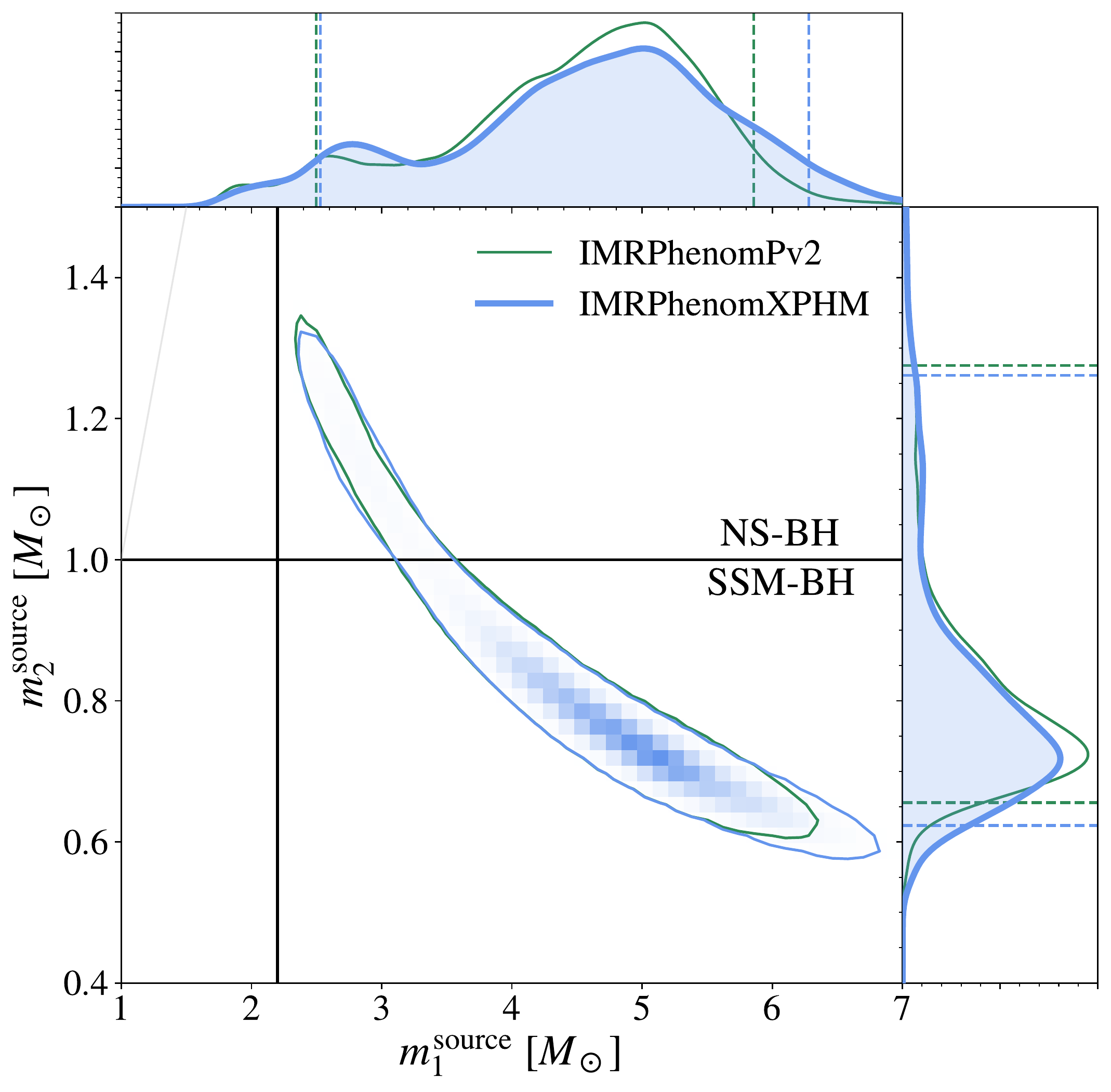}
\end{center} 
\caption{Posterior distributions for the primary and secondary mass in the source frame.  The 90$\%$ credible regions are indicated by the solid contour in the joint distribution, and by the dashed vertical and horizontal lines in the marginalized distributions. {We also paint a vertical line for the upper bound in the mass of any given Neutron Star and a horizontal one for the subsolar mass threshold.}}
\label{fig:lalinference_m1m2corner}
\end{figure}

We use a low-frequency cutoff of 20 Hz in both detectors for the likelihood evaluation and choose uninformative and wide priors. The primary tool used for sampling the posterior distribution is the \texttt{LALInference} Markov Chain Monte Carlo implementation as described in \cite{Veitchetal}. The power spectral density used in the calculations of the likelihood is estimated using \texttt{BayesWave} \cite{Cornish_2015,PhysRevD.91.084034}. The study uses the O2 open access data~\cite{Abbott:2019ebz} with a sampling frequency of 4096 Hz; however the likelihood is integrated up to 1600 Hz.

The best fit CBC template has $\sim$3000 cycles in the detector, allowing us to constrain with relatively high accuracy the source properties of SSM170401 in spite of the low SNR \cite{Cutler:1992tc}.  The estimated parameters are reported in Table~\ref{table:parameters}. The marginalized posterior for the absolute value of the matched filter SNR is \SNRmfPv~ for \texttt{IMRPhenomPv2} and  \SNRmfXPHM~ for \texttt{IMRPhenomXPHM}.
The median value of the SNR is lower than that found by the search, which was $8.67$. However, these two quantities are not directly comparable.  The SNR from the search is obtained by maximizing the ranking statistic over a discrete template bank \cite{Phukon:2021cus, messick2017analysis, cannon2012toward, CANNON2021100680}, while the quoted SNR from the PE is the median value over the samples.  Since the ranking statistic and the SNR are closely related, the SNR that is more comparable to that of the search would be the maximum SNR as found by the PE.  The values of this maximum PE SNR are \MaxSNRmfPv~ for \texttt{IMRPhenomPv2} and \MaxSNRmfXPHM~for \texttt{IMRPhenomXPHM}. These values are slightly larger than that of the search, which is consistent with what would happen if the signal was astrophysical.  However, this is also expected in the noise case due to the larger parameter space that allows more flexibility for the PE analysis to fit the data. We also notice the maximum value of the SNR to be larger for \texttt{IMRPhenomXPHM} than for \texttt{IMRPhenomPv2}. In a similar way, this is expected for an astrophysical signal but also for noise, since the waveform includes Higher Order Modes and thus has more flexibility to fit the data.

The {source} is then compatible with a compact binary system having an unequal mass ratio $q = $\qPv~ (all uncertainties are quoted at 90\% C.L.), a source frame primary mass $m_1 = \text{\PrimMassPv}M_{\odot}$ and a source frame secondary mass $m_2 = \text{\SecMassPv}M_{\odot}$ as shown in Fig.~\ref{fig:lalinference_m1m2corner}. The marginalised posterior distribution for the secondary mass favors a mass lower than $1 M_\odot$ (\ProbSSMPv~C.L.).  Using the \texttt{IMRPhenomXPHM} waveform, we find almost identical results, with a mass lower than $1 M_\odot$ at \ProbSSMXPHM~C.L. 

\begin{figure*}[t!]
\begin{center}
\includegraphics[width=0.48\textwidth]{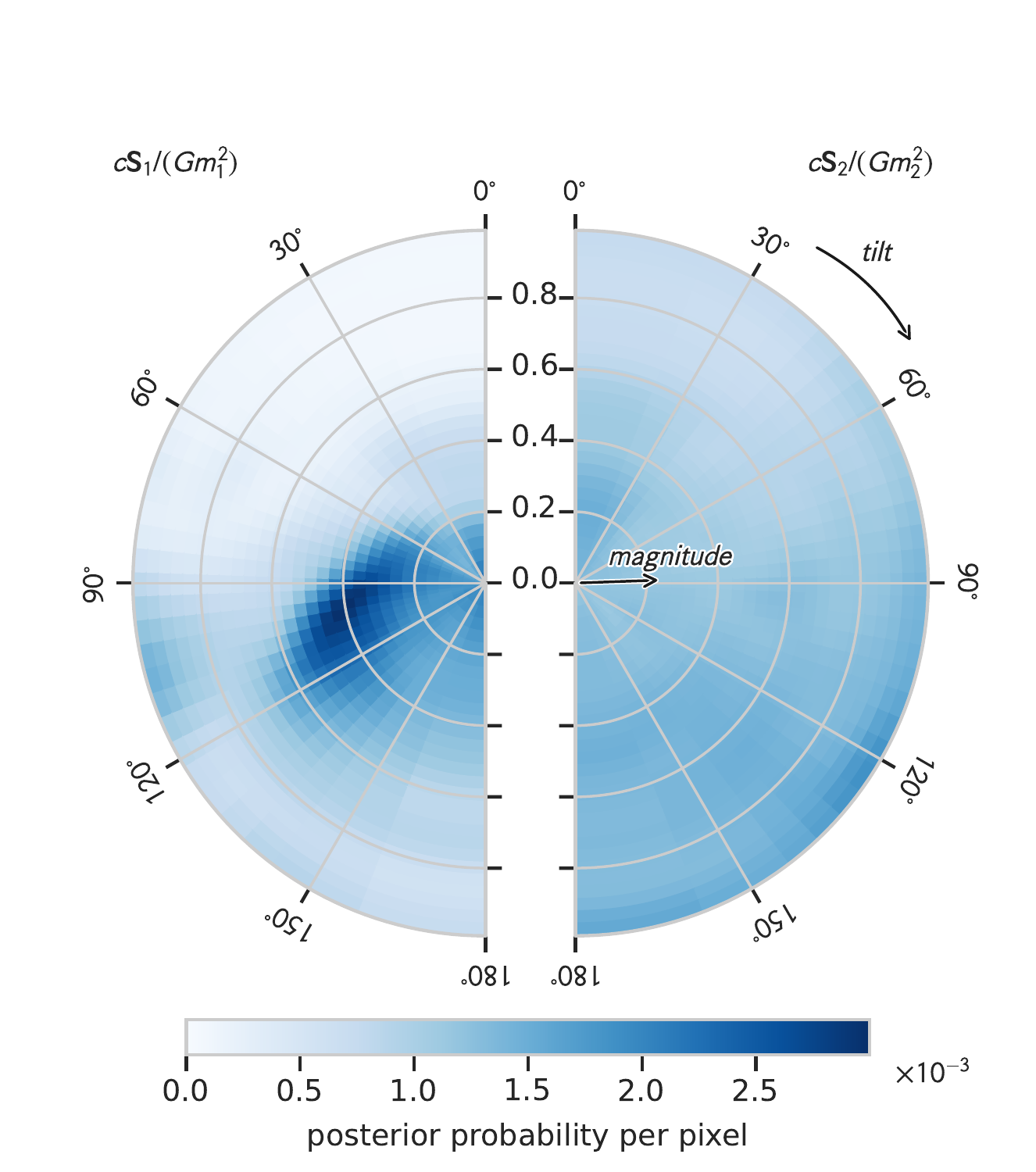}
\includegraphics[width=0.48\textwidth]{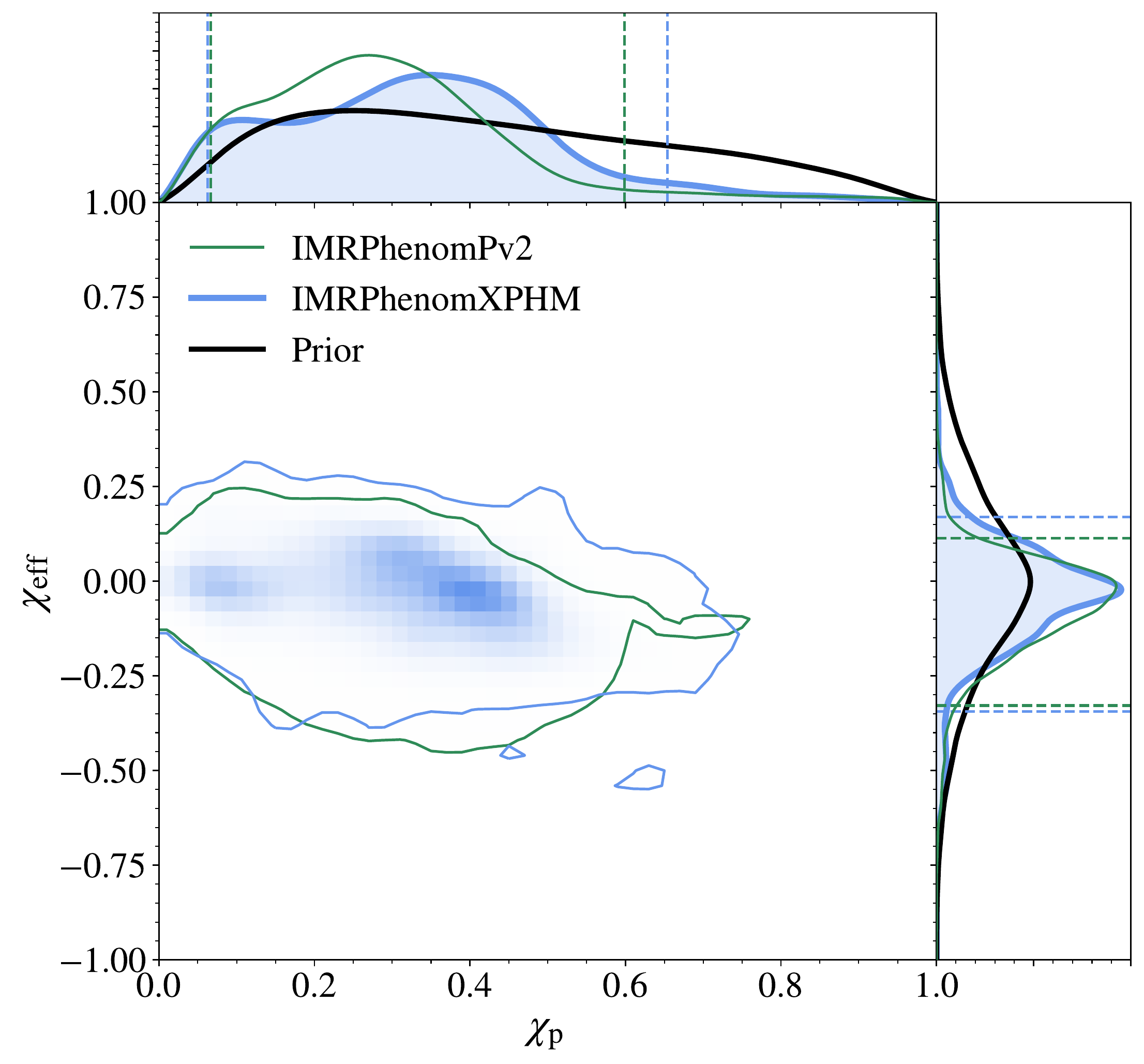}
\end{center} 
\caption{Left: posterior distribution for the individual spins of the source of SSM170401, according to the \texttt{IMRPhenomXPHM}  waveform models. The radial coordinate in the plot denotes the dimensionless spin magnitude, while the angle denotes the spin tilt, defined as the angle between the spin and the orbital angular momentum of the binary at a reference frequency of 20 Hz. A tilt of $0^\circ$ indicates that the spin is aligned with the orbital angular momentum. A nonzero magnitude and a tilt away from $0^\circ$ and $180^\circ$ imply a precessing orbital plane. All bins have an equal prior probability. Right: posterior distributions for the effective spin and effective in-plane spin parameters. The black lines in the right panel show the prior distributions for the effective spin parameters. The 90$\%$ credible regions are indicated by the solid contour in the joint distribution, and by dashed vertical and horizontal lines in the marginalized distributions. The large density for tilts close to $90^\circ$ leads to non-zero values for $\chi_{\text{p}}$ and low values for $\chi_{\text{eff}}$.
}
\label{fig:lalinference_Spins}
\end{figure*}

\begin{figure*}[t!]
\begin{center}
\includegraphics[width=0.45\textwidth]{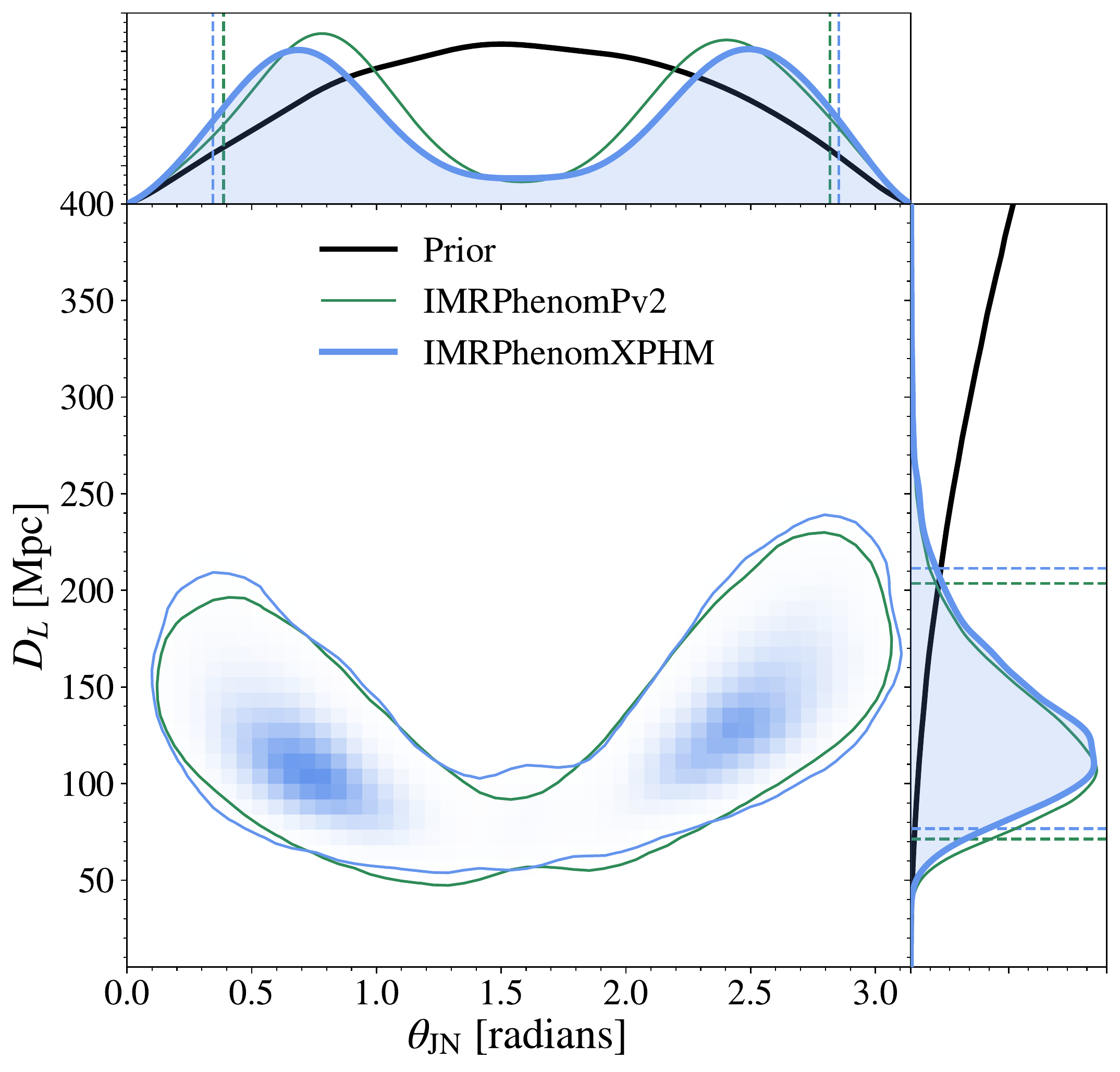}
\includegraphics[width=0.54\textwidth]{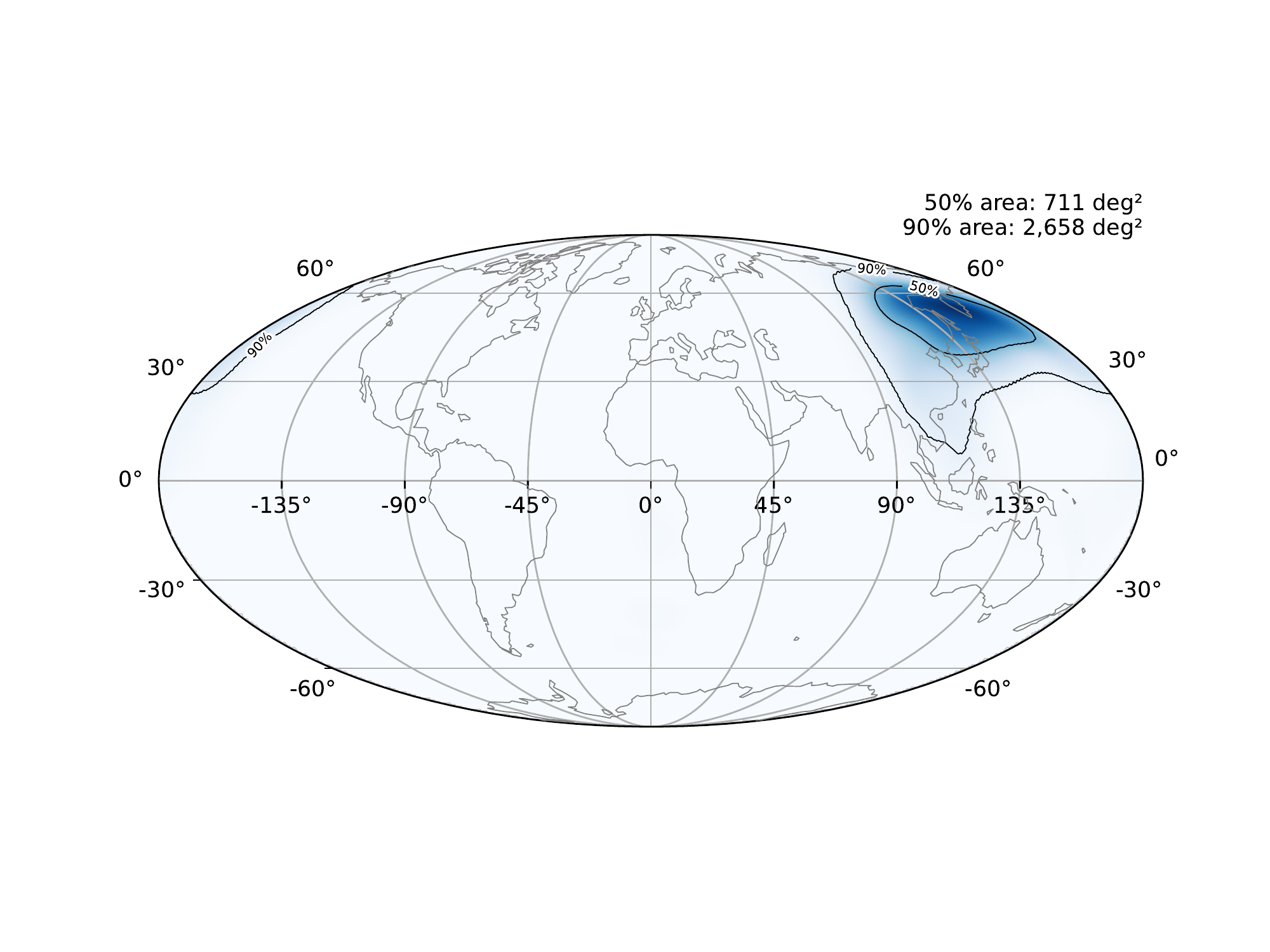}
\end{center} 
\caption{Left: posterior distributions for the luminosity distance and the inclination angle of the source of SSM170401, according to the \texttt{IMRPhenomXPHM} and \texttt{IMRPhenomPv2} waveform models. The inclination angle indicates the angle between the line of sight and the total angular momentum of the binary. For nonprecessing binaries, this is equal to the angle between the orbital angular momentum and the line of sight. The solid lines and the central contour denote 90$\%$ credible regions. Right: sky position of the event as evaluated from the Greenwich meridian according to the \texttt{IMRPhenomXPHM} waveform model.}
\label{fig:lalinference_corner_dL_inc}
\end{figure*}

The left panel of Fig.~\ref{fig:lalinference_Spins} shows the posterior distributions for the magnitude and tilt angle of the individual spins, measured at a reference frequency of 20~Hz. All pixels in this plot have an equal prior probability. The spin of the secondary BH is largely unconstrained, as expected for very unequal masses, {while the primary spin shows a preference for small spin magnitudes ($a_1=$\PrimSpinPv), where the posterior samples with large primary spin tend to have it misaligned with the orbital angular momentum.} As can be seen in the right panel of Fig.~\ref{fig:lalinference_Spins}, this leads to a $\chi_{\text{eff}}$ compatible with zero ($\chi_{\text{eff}}=$\ChieffXPHM) and an uninformative posterior in  $\chi_{p}$ ($\chi_p=$\ChipXPHM).

The luminosity distance and inclination angle $\theta_{J N}$ posterior distributions are shown together in the left panel of Fig.~\ref{fig:lalinference_corner_dL_inc}, since these two quantities are correlated.  We find a luminosity distance of $d_L=$\DLPv Mpc.  We identify a bimodal distribution for $\theta_{J N}$ due to the fact that we can not distinguish whether the system is being observed face-on ($\theta_{J N} \sim 0$) or face-away ($\theta_{J N} \sim \pi$), but it being edge-on ($\theta_{J N} \sim \pi/2$) is disfavoured. In the face-on(away) configuration, the effects of precession and higher order modes in the signal are suppressed \cite{Krishnendu:2021cyi,PhysRevD.95.104038,Mills:2020thr}, as is the case here.

\begin{figure}[t!]
\begin{center}
\includegraphics[width=0.48\textwidth]{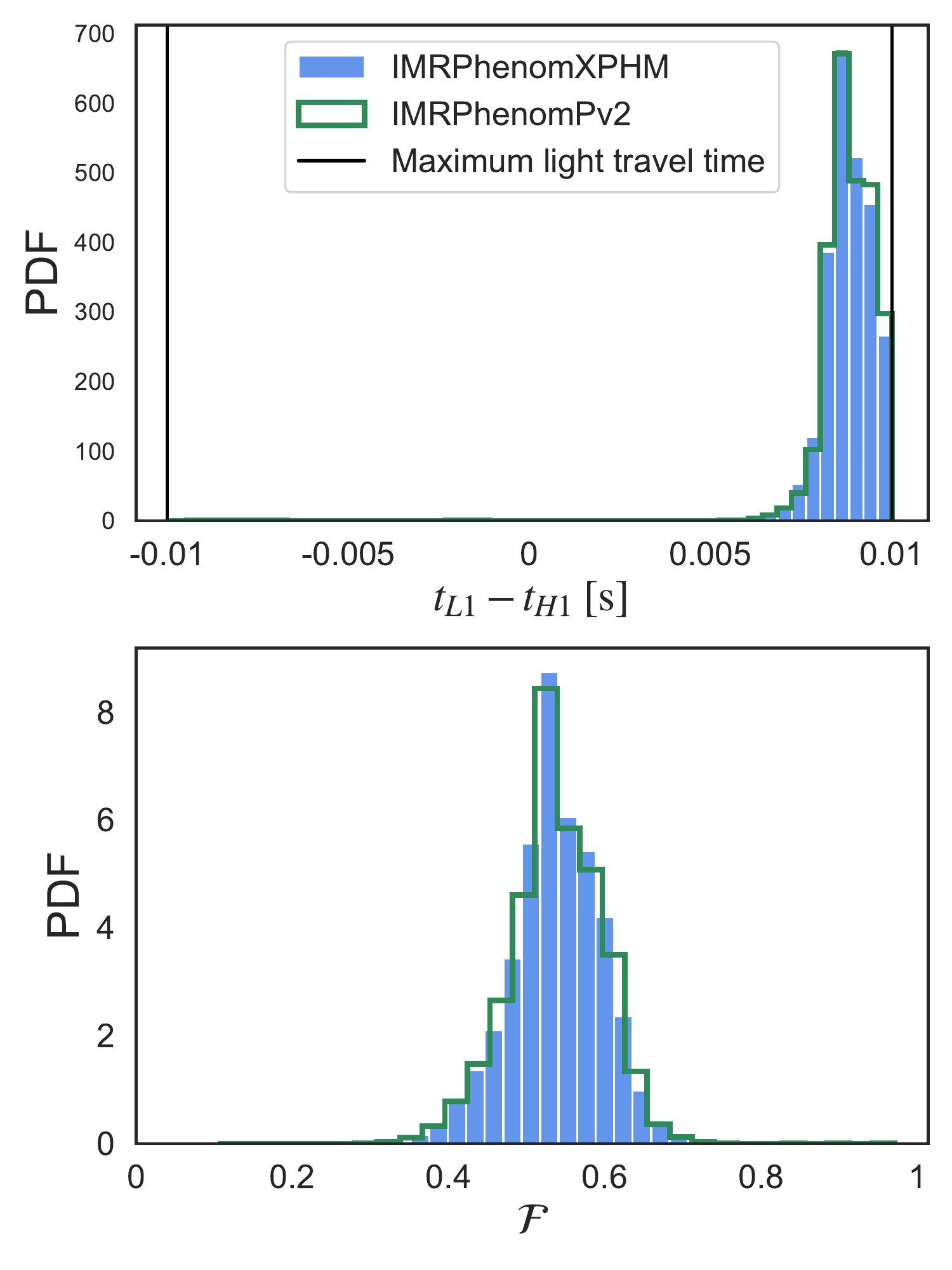}
\end{center} 
\caption{{Upper panel: posterior distribution of the time delay between the arrival of the signal in LIGO Livingston and LIGO Hanford. Lower panel: posterior distribution of the network antenna pattern $\mathcal{F}$, computed using Eq.~\eqref{eq:F_net}.}}
\label{fig:Dt_F_corner}
\end{figure}

{In the right panel of Fig.~\ref{fig:lalinference_corner_dL_inc}, we show the posterior distribution of the location in the sky of the event. This sky map looks abnormal when compared with the typical ones of the events detected exclusively by Hanford and Livingston~\cite{LIGOScientific:2018mvr, LIGOScientific:2020ibl, LIGOScientific:2021usb, LIGOScientific:2021djp}. When the trigger is seen in two detectors only, most of the information for the sky localisation comes from the time delay between the observation of the signal in both interferometers. This time delay will be given by:
\begin{equation}
    \Delta t_{L-H} = \frac{\Vec{d}_{H-L} \cdot \hat{n}}{c} = \frac{d_{H-L}}{c} \cos{\theta} \, ,
    \label{eq:t_delay}
\end{equation}
\noindent where $\Vec{d}_{H-L}$ is the position vector of Livingston with respect to Hanford and $\hat{n}$ is the direction in the sky of the source. We observe in Eq.~\eqref{eq:t_delay} that the time delay only constraints the inclination angle with respect to $\Vec{d}_{H-L}$, but leaves the azimuthal angle completely unconstrained. This has as a result a ring-like shape in the sky maps usually observed. However, when the source direction corresponds to $\theta = 0, \pi$, that is, the line joining both detectors, the ring will collapse to have a blob like shape in the sky. As can be seen in the top panel of Figure~\ref{fig:Dt_F_corner}, this is what is happening for the source of SSM170401, since the time delay between Livingston and Hanford is close to the maximum light travel time. In the bottom panel of Figure~\ref{fig:Dt_F_corner} we also show the posterior distribution of the network antenna pattern $\mathcal{F}$ \cite{Klimenko_2011}, defined as:
\begin{equation}
    \mathcal{F} = \sqrt{\frac{F_{+,\mathrm{H1}}^2 + F_{\times,\mathrm{H1}}^2 + F_{+,\mathrm{L1}}^2 + F_{\times,\mathrm{L1}}^2}{2}}
    \label{eq:F_net}
\end{equation}
\noindent where $F_{+(\times),D}$ are the $+$ ($\times$) antenna patterns of detector $D$.  In this plot we observe that the event is coming from a region in the sky where the network antenna pattern is significantly smaller than 1, peaking at $\mathcal{F} \sim 0.5$. This means that the sensitivity in this direction will be half of that of the most sensitive direction ($\mathcal{F} \sim 1$) located on top of the continental US and its antipodes~\cite{Andersson_2013}. Since the direction joining both LIGO detectors has smaller sensitivity, the distance up to which LIGO can detect astrophysical signals is also smaller. This leads to a lower expected event rate coming from that direction, which is the reason sky maps like the one of Fig.~\ref{fig:lalinference_corner_dL_inc} are uncommon.
}

Even though the posterior PDFs for the parameters of the source of SSM170401 seem to have converged to a well-defined distribution that differs from the prior, it is known that GW signals can be mimicked by gaussian noise~\cite{Morras:2022ysx} or non-gaussian transients, specially given the relatively low SNR and high FAR.

\subsection{Coherence Test}
\label{sec:properties_of_the_source:coherence_test}

\begin{figure}
\includegraphics[width=0.4\textwidth]{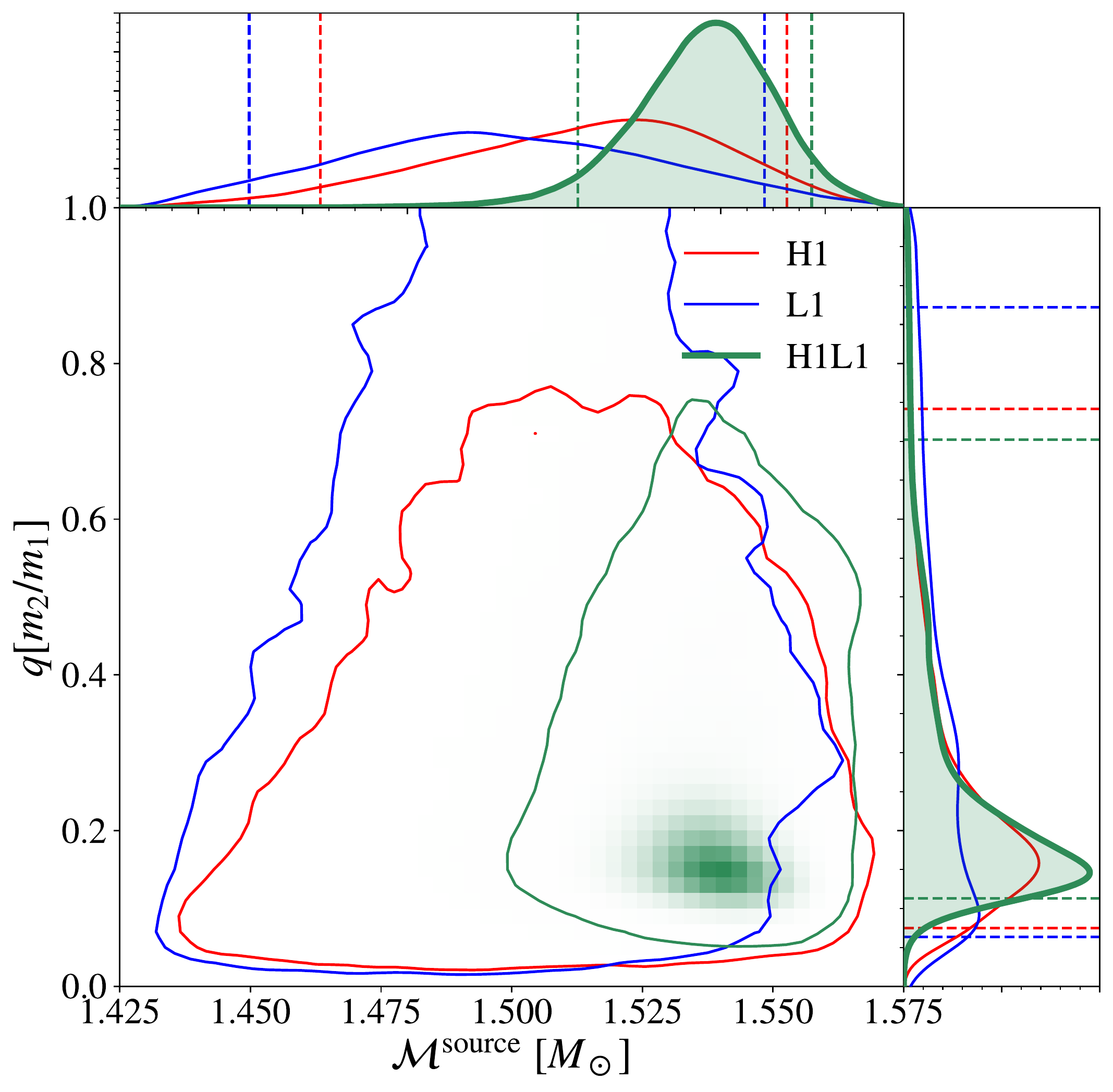}
\caption{{Posterior distributions for the mass ratio and the source frame chirp mass ($\mathcal{M}$) for the PE analysis performed with the waveform \texttt{IMRPhenomPv2} when considering both detectors together (H1L1) and individually. }}
\label{fig:coherence_Mc_q_corner}
\end{figure}

To test the compatibility of the SSM170401 with a GW coming from a CBC, we perform the coherence test proposed in Ref.~\cite{Veitch:2009hd}. The idea of this test is to perform Bayesian PE using the data from all the detectors together to calculate the evidence $Z_\mathrm{coh}$ for a coherent CBC signal and compare this with the evidence $Z_\mathrm{inc}$ for incoherent CBC signals. The incoherent evidence is defined as the product of the CBC signal evidences obtained performing PE individually in each detector. These incoherent CBC signals are used to represent noise in the detectors that can be picked up by CBC templates. The coherent versus incoherent hypothesis Bayes factor is then defined as

\begin{equation}
    \mathcal{B}_{\mathrm{coh},\mathrm{inc}} = \frac{Z_\mathrm{coh}}{Z_\mathrm{inc}} =  \frac{Z_\mathrm{coh}}{\prod_{i=1}^N Z^{(i)}} = \frac{\mathcal{B}_\mathrm{coh}}{\prod_{i=1}^N \mathcal{B}^{(i)}} \, ,
    \label{eq:Bcohinc}
\end{equation}

\noindent 
where we have used that for each interferometer, the Bayes factor of the signal versus noise hypothesis is defined as $\mathcal{B}^{(i)} = Z^{(i)}/Z^{(i)}_\mathrm{noise}$, while in the coherent analysis it is defined as $\mathcal{B}_\mathrm{coh} = Z_\mathrm{coh}/\prod_{i=1}^N Z^{(i)}_\mathrm{noise}$. We use the same priors for coherent and single-detector analyses. To get a more reliable estimate of the evidence, in the PE we use nested sampling, particularly the \texttt{Dynesty} sampler~\cite{speagle2020dynesty} as implemented in \texttt{Bilby}~\cite{Ashton:2018jfp}. The computational cost of performing the coherence test will be very large since it requires us to perform three separate PEs with the costly nested sampling. To make this analysis feasible we employ Reduced Order Quadrature (ROQ) methods~\cite{Smith:2016qas}, which greatly speed up the computation time of the likelihood, specially for long signals like SSM170401. The analysis is done using only the \texttt{IMRPhenomPv2} waveform, since we have seen that it gives consistent results with \texttt{IMRPhenomXPHM} while leading to much greater ROQ speedups~\cite{Qi:2020lfr}. We obtain the Bayes factors listed in Table.~\ref{table:coh_inc_evidences}. We find a value for $\log{\mathcal{B}_{\mathrm{coh},\mathrm{inc}}}=$\lnBcohinc, strongly favoring the coherent hypothesis over the incoherent hypothesis~\cite{Trotta:2008qt}.

This result, however, cannot be used to update the statistical significance of the candidate since we have not run the coherence test over the background triggers of the search. While it is unlikely that a randomly selected noise candidate would give such a large value of $\log{\mathcal{B}_{\mathrm{coh},\mathrm{inc}}}$~\cite{Isi:2018vst}, we note that the search ranking promotes candidates with parameters consistent between different detectors~\cite{Cannon:2015gha}, thus it may be less surprising that a highly ranked candidate has large $\log{\mathcal{B}_{\mathrm{coh},\mathrm{inc}}}$. In Fig.~\ref{fig:coherence_Mc_q_corner} we also show the posterior distributions of $(\mathcal{M}, q)$ obtained performing PE in each detector individually and coherently in both of them. We observe that the 2D contours are compatible with each other, having larger areas in the single-detector analyses. This behavior is what is expected if the signal in both detectors were generated by the GWs coming from a single CBC~\cite{Veitch:2009hd}.

\begin{table}[t!]
    \bgroup
    \def\arraystretch{1.5}
	\begin{tabular}{| c | c | c | c |}
        \hline
    	$\log{\mathcal{B}_{H1L1}}$ & $\log{\mathcal{B}_{H1}}$ & $\log{\mathcal{B}_{L1}}$ & $\log{\mathcal{B}_{\mathrm{coh},\mathrm{inc}}}$ \\ 
        \hline
        \lnBHL & \lnBH & \lnBL & \lnBcohinc \\
        \hline
    \end{tabular}
    \egroup
    \caption{{Natural logarithm of the Bayes factors of the signal versus noise hypotheses obtained from the PE in the data of Hanford-Livingstion $\log{\mathcal{B}_{H1L1}}$, only Hanford ($\log{\mathcal{B}_{H1}}$), only Livingston ($\log{\mathcal{B}_{L1}}$) and the natural logarithm of the Bayes factor of the coherent versus incoherent hypothesis $\log{\mathcal{B}_{\mathrm{coh},\mathrm{inc}}} = \log{\mathcal{B}_{H1L1}} - \log{\mathcal{B}_{H1}} - \log{\mathcal{B}_{L1}}$.}}
\label{table:coh_inc_evidences}
\end{table}

\section{Discussion}
 
To discuss the possible source of SSM170401, assuming it is a CBC, we have divided the $(m_1,m_2)$ in four regions, according to the SSM threshold ($m_2 = 1~M_\odot$) and the maximum allowed mass of a NS ($m_1 = 2.2~M_\odot$)~\cite{Margalit:2017dij,Ruiz:2017due,NANOGrav:2019jur}. We observe that the full $90\%$ credible region of the posterior lies in the region of $m_1 > 2.2~M_\odot$, excluding the NS origin of the primary component. We find that $16\%$ of the posterior distribution lies in the region of $m_2 > 1~M_\odot$, which would point to a likely NS origin, although a light black hole cannot be excluded. However, the most probable region, representing \ProbSSMXPHM~of the posterior, would imply a mass of the secondary component below $1~M_\odot$. It is thus interesting to explore what could be the origin and nature of a possible SSM object.

{The first possibility to consider for an SSM component would be a neutron star.} Neutron stars have relatively well-determined masses from observations of binary systems, including pulsars or X-ray binaries involving an accreting neutron star from a companion.  Their {measured} masses are {above $1.2~\Msun$~\cite{universe5070159}}, further confirmed by the observation of GW170817~\cite{TheLIGOScientific:2017qsa}. {However, there is a recent claim~\cite{2022NatAs.tmp..224D} for a neutron star of mass $0.77^{+0.20}_{-0.17} M_\odot$, although it has been argued~\cite{Sotani:2013dga} that such a small mass for a neutron star probably requires a strange QCD equation of state. Therefore, the neutron star interpretation of a possible SSM component cannot be excluded, although it is disfavored by the bulk of observational data.}

{Another possibility is} PBHs formed by the gravitational collapse of large inhomogeneities in the early Universe {which} are already considered as a possible explanation of LVK GW detections, see e.g.~\cite{Bird:2016dcv,Clesse:2016vqa,Sasaki:2016jop,Clesse:2017bsw,Fernandez:2019kyb,Carr:2019kxo,Jedamzik:2020ypm,Jedamzik:2020omx,Clesse:2020ghq,Fernandez:2019kyb,Hutsi:2020sol,Hall:2020daa,Garcia-Bellido:2020pwq,Franciolini:2021tla}.  Depending on the model, they may explain anything from a tiny fraction of Dark Matter to its entirety. PBHs have been the main motivation to conduct searches of SSM black holes in the LVK data~\cite{Abbott:2018oah,abbott2019search,Nitz:2020bdb,Phukon:2021cus,Nitz:2021mzz,LIGOScientific:2021job,Nitz:2022ltl}, in particular, the extended subsolar search with low-mass ratios in O2
which reported SSM170401 as a possible candidate~\cite{Phukon:2021cus}. If some of the observed binary coalescences are indeed due to PBHs, they must have a relatively extended mass distribution that would have been imprinted by the thermal history of the Universe~\cite{Byrnes:2018clq,Carr:2019kxo}.  
This would lead to a peak in the mass distribution around a solar mass which is naturally produced at the QCD transition~\cite{Jedamzik:1996mr,Niemeyer:1997mt,Byrnes:2018clq,Carr:2019kxo,Garcia-Bellido:2021zgu,Escriva:2022bwe,Franciolini:2022tfm}, and the source of SSM170401 could be an example of a subsolar PBH around the QCD-induced peak. 
{The spin posterior is quite broad and the spin is compatible with zero, although a slight preference for a primary spin around 0.3 is observed.}
In this case, the non-zero but relatively low spin of the primary component may have been acquired by matter accretion, previous mergers or hyperbolic encounters~\cite{DeLuca:2020bjf,Garcia-Bellido:2020pwq,Jaraba:2021ces}.  

{Alternatively, in scenarios with complex and dissipative particle Dark Matter, SSM black holes could form through the cooling and gravitational collapse of Dark Matter halos~\cite{Shandera:2018xkn}. This model was constrained by the LVK data in~\cite{Singh:2020wiq,LIGOScientific:2021job,LVK:O3bSSM}. Furthermore, it could be that the secondary component of the source of SSM170401 is a boson star, a hypothetical horizonless compact object formed by an ultralight bosonic field. If the mass of the bosonic particle is larger than $10^{-10} {\rm eV}/c^2 $, the boson star can have subsolar mass~\cite{Liebling:2012fv}. Whether a merger with an SSM boson star component could produce a signal similar to the SSM170401 trigger, though, remains to be investigated.}

{Finally, we note that the primary component mass of the hypothetical source of SSM170401 would preferably lie in the hypothesized low mass gap between $2.5$ and $5 M_\odot$ (\ProbMGPv C.L.). However, this is not unique, since other candidates with components possibly in this lower mass gap have been observed, namely GW190814~\cite{Abbott:2020khf} and GW200210$_{-}$092254~\cite{LIGOScientific:2021djp}.}

\section{Conclusions}

{We have performed an in-depth investigation of the most significant double-detector candidate reported in~\cite{Phukon:2021cus} in an SSM search over O2 data. We have removed a prominent blip glitch 14 s before coalescence in the data and estimated the parameters of the possible CBC source using a low frequency cutoff of 20 Hz. Parameter Estimation runs were performed using \texttt{LALInference} with two different waveforms \texttt{IMRPhenomPv2} and \texttt{IMRPhenomXPHM}, where the latter includes effects from higher order modes. The source parameters obtained by both PE runs show good agreement with each other and with the parameters of the template that triggered the search. We find a median network SNR of \SNRmfXPHM ($90\%$ credible interval), which is lower than the SNR of $8.6$ obtained in the search. However, the search SNR is more closely related to the maximum SNR, which we find to be higher in the PE, where it reaches values of \MaxSNRmfPv~ for \texttt{IMRPhenomPv2} and \MaxSNRmfXPHM~for \texttt{IMRPhenomXPHM}. The secondary mass is $m_2 = \text{\SecMassXPHM}M_\odot$ ($90\%$ credible interval), with \ProbSSMXPHM~confidence of being below one solar mass. For the location in the sky posterior, we find an atypical distribution when compared with the usual Hanford-Livingston events detected thus far, which can be explained if it were a GW coming from the direction joining the two LIGO interferometers.}

{The compatibility of SSM170401 with a CBC origin has been further tested by performing the signal coherence tests of Ref~\cite{Veitch:2009hd}, obtaining a log Bayes factor of \lnBcohinc~for the coherent vs incoherent hypothesis. Furthermore, we observe that the $(\mathcal{M}, q)$ posteriors of each independent IFO converge to mutually compatible contours. These tests provide significant support in favor of a coherent signal, which generally is not expected if it were generated by noise fluctuations~\cite{Isi:2018vst}. We also checked that the O3 limits on the SSM merger rate~\cite{LVK:O3bSSM} do not put a significant constraint on the probability of this candidate being astrophysical ($P_{90} \lesssim 0.45$). Therefore, we do not find compelling arguments against a possible CBC origin of SSM170401.}

{Finally, even if most of the $m_2$ posterior support is in the SSM region, there is still a $16\%$ probability of $m_2$ being over 1 $\Msun$, which does not allow us to convincingly exclude a NS origin. Candidates with a higher SNR would have better measurements on their parameters, allowing for more confident discrimination between sub- and super- solar masses~\cite{Wolfe:2023yuu}. Therefore, the data from future planned LIGO-Virgo-Kagra runs with improved sensitivity~\cite{LVKarticle}, O4 and O5, offer a great opportunity for discovering CBC mergers with SSM components, if they are out there in the Cosmos.}

\acknowledgments{The authors thank Patrick Meyers and Walter Del Pozzo for their helpful comments and discussions as reviewers of this paper in LIGO and Virgo respectively as well as Juan Calderón Bustillo and Geraint Pratten for their constructive feedback. {We thank the anonymous referee for their careful reading of our manuscript and their many insightful comments and suggestions.} This work is partially  supported   by  the Spanish grants PID2020-113701GB-I00, PID2021-123012NB-C43 [MICINN-FEDER], and the Centro de Excelencia Severo Ochoa Program CEX2020-001007-S through IFT, some of which include ERDF  funds  from  the  European  Union. S.C. acknowledges support from the Francqui Foundation through a Starting Grant.  K.M. is supported by King's College London through a Postgraduate International Scholarship. M.S. is supported in part by the Science and Technology Facility Council (STFC), United Kingdom, under the research grant ST/P000258/1. IFAE  is  partially funded by the CERCA program of the Generalitat de Catalunya. We acknowledge the use of IUCAA LDG cluster Sarathi for the
computational/numerical work. This material is based upon work supported by NSF's LIGO Laboratory which is a major facility fully funded by the National Science Foundation.}

\bibliographystyle{apsrev4-1}

\bibliography{Biblio} 

\end{document}